\pdfoutput=1

\documentclass[subscriptcorrection,upint,varvw,mathalfa=cal=euler,balance,hyphenate,french,pdf-a,nofoot]{asmejour} %

\hypersetup{%
	pdfauthor={John H. Lienhard},                       		   	
	pdftitle={ASME Journal Paper LaTeX Template},                  	
	pdfkeywords={ASME journal paper, LaTeX template, BibTeX style, asmejour class},
	pdfsubject = {Describes the asmejour LaTeX template},			
	pdflicenseurl={https://ctan.org/pkg/asmejour},
}

\usepackage{etoolbox}
\AtBeginEnvironment{quote}{\par\singlespacing\small}
\usepackage{setspace}
\usepackage{dblfloatfix}
                   
\begin{document}

\SetAuthorBlock{Garrett Russell}{
   Massachusetts, U.S.A., \\
   email: garrettrussell111@gmail.com} 

\title{The Properties of Glass Fiber Reinforced Polypropylene Filaments Recycled from Fishing Gear}


\begin{abstract}
Plastic pollution, mainly from lost fishing gear composed of high-density polyethylene (HDPE) and polypropylene (PP), poses a significant environmental obstacle. This study evaluates the potential of recycling PP from fishnet/rope and reinforcing it with glass fiber (GF) in the form of 3D printer filaments as a way to reduce/prevent ocean plastic. Two materials, one virgin (vPP-GF) and one made up of recycled polypropylene and virgin glass fibers (rPP-GF), were analyzed using differential scanning calorimetry, tensile, and Charpy impact tests. From the results, it was found that rPP-GF often outperformed vPP-GF. rPP-GF had a higher melting and crystallization point, likely a higher crystallinity, and could withstand a higher tensile stress, while vPP-GF could withstand a higher tensile strain. Further analysis revealed the potential presence of HDPE within the rPP-GF composite, which was not reported by the manufacturer. This significantly affected the Charpy test and made it difficult to draw conclusions from the resulting data. Nevertheless, their comparability in terms of mechanical and material properties indicates the strong potential of recycling polypropylene fishnet/rope and reinforcing it with glass fibers to extend their lifespan and reduce ocean plastic. 
\vspace{.7em} 
\end{abstract}

\date{}

\maketitle

\section{Introduction}

In recent years, plastic pollution, especially in the ocean, has come to the forefront of pressing environmental issues. Approximately 100,000 tons of harmful\footnote{Plastic pollution has several negative environmental effects: harming marine life, marine eco-systems, and even socio-economic areas like fishing or human health \cite{Thushari2020}.} plastic have accumulated in our oceans since 1950, nearly a third of which is composed of high-density polyethylene (HDPE) and polypropylene (PP) \cite{Maldonado-García2021}. Notably, 75 to 86\verb|%| of this plastic originates from lost fishing gear, according to a \textit{Nature} published study on the North Pacific Garbage Patch \cite{Lebreton2022}. This damaging environmental impression from human practice evidently necessitates the development of innovative solutions to cleanse the oceans. Therefore, the ultimate goal of this research is to explore the viability of recycling plastic to prevent and/or reduce ocean plastic.

Recent developments in the 3D printing industry, which relies heavily on plastic use, include commercially available desktop printers, novel infill geometries\footnote{An infill is the inside structure of a printed part. Some researchers have chosen to investigate unique versions, such as geometry inspired by origami \cite{Shen2022}.}, and even the introduction of unique methods of printing (like metal printing or nanoprinting). This, on top of the numerous applications, such as in the aerospace and automotive industries, has fueled exponential growth  \cite{Montez2022}. It reached a market value of \$12.6 billion in 2020 and is expected to increase to \verb|$|37.2 billion by 2026 \cite{Placek2022}. Given that this will only result in an increased use of plastics, a particularly promising way of reducing the patch arises: recycling fishing gear in the form of filaments\footnote{A filament is a small string of plastic that can be fed into a 3D printer in order to extrude objects.} to be printed and sold. Recycling non-ocean plastic into filament has already been demonstrated to be possible with other materials like polylactic acid (PLA) and polyethylene terephthalate glycol (PET-G).

This is a possible solution because both PP and HDPE are thermoplastic polymers, a type of material that is commonly used in the 3D printing industry. HDPE, however, is difficult to print because of its high tendency to warp\footnote{Following an object being printed, it can deviate from the intended shape if the part does not shrink uniformly \cite{Schmutzler2016}.} and consequently is scarcely available in filament form online. PP is widely used (both in and outside of 3D printing), mainly for its corrosion resistance, low cost, and simple processing, though it is limited in its engineering applications because of its low mechanical properties \cite{LUO2018198}. PP also suffers from a high warpage, limiting its applications to expert printers \cite{prusa}. On the other hand, reinforcing PP with a glass fiber (PP-GF) reduces warpage \cite{Sikló2011} and increases the mechanical strength of the polymer. Despite still being difficult to print, its corrosion and thermal resistance, strength, and low cost have opened up multiple industrial applications for the material, including pipes, automobiles, and household appliances. \cite{Luo2017}. Considering HDPE is currently unavailable for purchase, only PP-GF and PP were chosen to be further investigated in this review.

The central concept behind 3D printing involves the construction of objects layer by layer, and while there are various processes to achieve this, only fused filament fabrication and fused deposition modeling were considered in this study, mainly for accessibility reasons. Both methods are similar and involve the use of a nozzle to extrude the fibers of a material into a design \cite{Dezfooli2020}.


\begin{table*}[!b]


\caption{Results of studies comparing recycled PP (rPP) from waste/used material with virgin PP (vPP)}\label{tab:0}%

\centering{%
\begin{tabular*}{0.8\textwidth}{@{\hspace*{1.5em}}@{\extracolsep{\fill}}p{7em}p{7em}p{16em}p{5em}}
\toprule

\textbf{Material} & 
\textbf{Origin} & 
\textbf{Result} & 
\textbf{Reference} \\

\midrule
\addlinespace
varying vPP/rPP blends & N/A & higher recycled=slightly lower tensile strength & \cite{Jamnongkan2022} \\
\addlinespace
varying vPP/rPP blends & municipal \newline collection centre & blend degrades with increased recycled material & \cite{Raj2013} \\
\addlinespace
ocean rPP and \newline vPP & marine plastic & polymer mechanical properties degrade & \cite{Pelegrini2019} \\
\addlinespace
rPP & post-consumer & lower tensile and impact strength \newline (material becomes more brittle) & \cite{Satya2020} \\
\addlinespace
\bottomrule
\addlinespace[1.2em]
\end{tabular*}
}%


\caption{Results of studies investigating the effects of recycling vPP}\label{tab:1}%

\centering{%
\begin{tabular*}{0.8\textwidth}{@{\hspace*{1.5em}}@{\extracolsep{\fill}}p{8em}p{18em}p{6em}}
\toprule

\textbf{Material} & 
\textbf{Result} & 
\textbf{Reference} \\

\midrule
\addlinespace
vPP and rPP & higher crystallinity, elastic modulus, yield stress; material becomes more brittle & \cite{Aurrekoetxea2001} \\
\addlinespace 
varying vPP/rPP \newline blends & positive impact on some tensile properties, opposite for flexural & \cite{Hyie2019} \\
\addlinespace 
vPP and rPP & increase in tensile strength, crystallinity & \cite{Huang2021} \\ 
\addlinespace
vPP and rPP & decrease in elongation at break & \cite{Incarnato1999} \\
\addlinespace

\bottomrule
\addlinespace[1.2em]
\end{tabular*}
}%


\caption{Results of studies investigating PP-GF \\ (the number following indicates the wt\% of GF)}\label{tab:2}%

\centering{%
\begin{tabular*}{0.8\textwidth}{@{\hspace*{1.5em}}@{\extracolsep{\fill}}p{7em}p{7em}p{16em}p{5em}}
\toprule

\textbf{Material} & 
\textbf{Origin} & 
\textbf{Result} & 
\textbf{Reference} \\

\midrule
\addlinespace
varying \newline rPP-GF/vPP blends & Bulk trim of \newline truck trailers & higher flexural strength & \cite{Vaidya2023} \\
\addlinespace 
vPP-GF40 & N/A & slight decrease in elastic modulus, \newline flexural and tensile strength; overall no significant effect & \cite{Colucci2015} \\
\addlinespace
vPP-GF(?) & N/A & decreased crystallinity, flexural \newline properties & \cite{Achukwu2023} \\
\addlinespace
vPP-GF10, \newline rPP-GF10 & PP \newline (industrial scrap) & properties diminish with recycling \newline (including crystallinity) & \cite{Shuib2023} \\
\addlinespace
vPP-GF30 & N/A & tensile strength degrades & \cite{Rafiq2013} \\
\addlinespace
vPP-GF50 & N/A & tensile strength and modulus degrades & \cite{Chen2020} \\
\addlinespace

\bottomrule
\addlinespace[1.2em]
\end{tabular*}
}%

\end{table*}
\section{Literature Review}


\subsection{Background on Polymer Properties}

\begin{figure}[ht]
\centering\includegraphics[width=.7\linewidth]{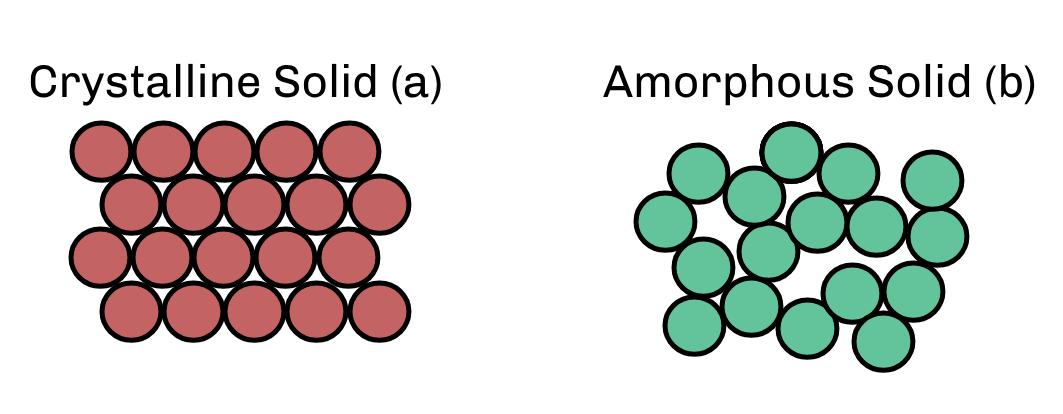}
\caption{An illustration of two examples of structures within a solid, one being highly ordered, or crystalline (a), and the other being highly disordered, or amorphous (b).\label{fig:1}}
\end{figure}

In regards to recycling, the properties of polymers can dramatically change \cite{Gomes2022}, so it is important to first understand what they are. When a material is repossessed, its crystallinity has been demonstrated to be a good indicator of changes in mechanical properties, or the characteristics of a material in response to an applied force \cite{Balani2014, Aurrekoetxea2001, Yasin2020}. Crystallinity is essentially the degree of order within a material \cite{Crawford2017}. If a material is highly ordered, it can be referred to as crystalline, and if it is highly disordered, it can be referred to as amorphous. See Fig.~\ref{fig:1} for a visual comparison between the two. 

\noindent Generally speaking, a crystalline polymer should have a higher tensile strength, while an amorphous polymer should have a higher impact strength\footnote{The tensile and impact strength are both properties determined by mechanical tests. See methods section for more details.} \cite{Balani2014,Odrobina2020}. There are many other aspects of a material that can change in response to recycling, however, for the purposes of this study, only the impact, tensile, and crystallinity properties were considered.

\subsection{A Review On PP and PP-GF Research}

To gauge the effectiveness of PP and PP-GF, a review of their material (crystallinity) and mechanical properties in response to recycling was conducted. It is relevant to survey both materials as PP is a component of PP-GF, which was ultimately determined to be the focus of the study. Given the above information, if an article did not explicitly state changes in crystallinity but did state changes in tensile or impact strength, the tensile and/or impact strength was taken as an indication of changes in crystallinity and vice versa. The materials were investigated separately and PP was reviewed first. For PP, the studies were then further separated into two categories: those which obtained previously recycled/used material and those that obtained a virgin material, the results of which can be found in Table \ref{tab:0} and Table \ref{tab:1}, respectively. In the case of PP-GF, most studies seemed to be of the same consensus and there were fewer in comparison to PP, so they were not separated based on the initial material. The results of this can be found in Table \ref{tab:2}. All studies reviewed employed mechanical recycling in order to reprocess their material.

\subsection{Discussion of Results of Literature Review}

\subsubsection{PP (waste/used)} In Table \ref{tab:0}, no studies reviewed found that recycling waste/used PP resulted in a positive impact on the mechanical properties or an increase in crystallinity, indicating that the materials may have experienced additional degradation as a result of the environment they came from. This would further separate their properties from vPP. Kauê Pelegrini et al., who investigated the degradation of PP and polyethylene from the marine environment, found that the materials experienced a variety of degradation mechanisms, which aligns with this conclusion. However, they also found that the materials can still be recycled and exhibit characteristics similar to other virgin materials on the market. Therefore, it seems likely that recycled PP ocean plastic will exhibit inferior mechanical properties when compared with virgin materials, but not to the extent that they are not comparable.

\subsubsection{vPP} An important distinction must be made between recycling virgin and waste/used materials. As virgin materials have yet to experience environmental degradation, a few of their properties (following reprocessing) seem to actually be impacted positively. This is demonstrated by the results of Table \ref{tab:1}. All studies reviewed found either an increase in tensile strength, brittleness, or crystallinity, indicating that recycling vPP likely increases crystallinity. J. Aurrekoetxea et al., who studied how the properties of vPP are affected by recycling, explains a possible reason for this quite nicely: 

\begin{quote}
“The observed reduction of the molecular weight with recycling increases the mobility and the ability to fold of the chains, allowing the formation of thicker lamellae and higher degree of crystallinity” \cite{Aurrekoetxea2001}.
\\
\end{quote}

\noindent Essentially, recycling the polymer causes degradation in the structure of the polymer itself, which allow the polymer to fold and form thicker lamellae, or regions of order within a material, and obtain a higher degree of crystallinity. If this is the case, it is likely recycling vPP will increase tensile strength and crystallinity while decreasing impact strength.

\subsubsection{PP-GF} Contrary to vPP, the majority of research reviewed on recycling PP-GF came to the conclusion that the properties degrade, even when beginning with vPP-GF \cite{Colucci2015,Achukwu2023,Rafiq2013,Chen2020}. The reason for this lies within the glass fibers. According to a literature review conducted by A. Pegoretti, an expert in polymers with over 30 years of experience as a professor at the University of Trento, it is common for both the fiber length and molecular weight to decrease after reprocessing, consequently reducing the mechanical properties \cite{PEGORETTI202193}. While a reduction in molecular weight allowed vPP to increase in crystallinity and tensile strength, both shortening the fiber length and reducing molecular weight simply degrades PP-GF.

\subsection{Gap Justification}

Researchers have yet to explore commercially available filaments (to the general consumer) and adding vGF as a reinforcement to rPP. It is this very disparity that the central question of this research arises: How do the properties of commercially available rPP filaments from ocean plastic and/or related material reinforced with vGF compare with pure vPP-GF filaments? This study will attempt to answer this question and take an approach different from anything previously surveyed. As evidenced by Tables \ref{tab:0}, \ref{tab:1}, and \ref{tab:2}, few simultaneously analyze the crystallinity and tensile or impact properties, and even fewer use an already used/recycled material. Only one, authored by Kauê Pelegrini et al., even places a specific emphasis on PP and the oceans \cite{Pelegrini2019}. Thus, the goal of this research is to bridge this gap, to investigate and add to the understanding of how PP responds to recycling and the addition of a GF, as well as explore a potential avenue of treating the world’s plastic pollution. By exploring all three properties (crystallinity, impact, tensile), this research can determine whether reinforcing rPP with vGF is a viable solution to treating/preventing plastic pollution. Based on what was found, it is predicted that a blend of waste/used rPP with vGF will exhibit degraded properties in comparison to vPP-GF. However, regardless of what is found, the results of this study will help guide and hopefully encourage others to further dive into ways to re-use materials and heal the environment.

\section{Methods}

\subsection{Materials}

To satisfy the goal of this study, there were a few requirements when it came to choosing the filaments to study. Namely, to reduce time, cost, and difficulty in obtaining the materials (for this research and others), the filaments had to be commercially available for purchase online; the weight percentage (wt\%)\footnote{In this case, wt\% refers to the weight percentage of PP within PP-GF. When comparing recycled and un-recycled materials, ensuring that their wt\% is similar is important to mitigate the possibility of differences in material composition affecting results.} of PP had to be within five percent of PP-GF30\footnote{PP-GF30 was selected for its prevalence online.}; and lastly, there had to be at least one recycled and un-recycled filament. This process yielded three potential options: vPP-GF30 from Ultrafuse, vPP-GF30 from 3DXTECH, and rPP-GF (recycled from fishnet and ropes) from Reflow. Ultrafuse was eliminated for budget reasons, so Reflow and 3DXTECH were decided to be the focus of this investigation. It is important to note that the rPP-GF does not originate directly from the oceans; rather, it originates from fishnet and rope at the end of their lifespan. Considering plastic in the ocean experiences additional degradation from the environment, the applications of this study to recycling material from the ocean may be limited. Regardless, recycling material (like fishnet and rope) to prevent it from reaching the ocean is equally as important as cleaning up the ocean itself. Therefore, investigating the two materials still holds paramount significance. A better alternative would be to create filaments from ocean plastic and compare them with virgin versions, but this would require extensive resources that would be beyond the budget and time constraints of this study.

\subsubsection{Important Material Notes}

In addition to the properties discussed above, it’s important to note some additional considerations regarding the use of the materials selected. Firstly, these materials are extremely difficult to print, as they warp frequently, have trouble adhering to the bed, and are sensitive to changes in printing parameters. Therefore, it is important that one uses the proper printing parameters, the proper bed adhesive, and a quality printer. Although print profiles do exist for PP-GF, it was found that the parameters had to be fine tuned through trial and error to obtain the best results. In terms of safety, volatile organic compounds (VOCs)\footnote{VOCs are ultrafine particles that are small enough to reach deep into the respiratory system. They can also be more difficult to clear \cite{Byrley2019}.} are emitted during the 3D printing process, which can be hazardous to human health \cite{Byrley2019}. To appropriately address this risk, a filter was used at all times while printing.

\subsection{Assessment Techniques}

The tensile, Charpy impact, and differential scanning calorimetry (DSC) tests were chosen to quantitatively analyze these materials through an experiment. ASTM\footnote{ASTM standards detail the process of designing, manufacturing, conditioning, testing, and analyzing specimens. They are commonly used to investigate materials, especially for engineering applications.} standards were used for the tensile and Charpy tests.

\subsubsection{Tensile Testing}

For its prevalence in the literature, the tensile test (ASTM D638-14) was chosen. In this test, the stress and strain of a dog-bone-shaped specimen are measured as it is pulled from two opposite ends until it breaks. Stress is defined as the force on the material divided by its cross-sectional area, while strain is the change in length of the body divided by the original length. Plotting stress versus strain via graphs is an exceptional technique for analyzing a material’s strength, stiffness, and ductility (flexibility), which is why it was employed in this study.

\subsubsection{Charpy Impact Testing}

The Charpy test (ASTM D6110-18) was chosen to (1) investigate the mechanical properties of the materials in a different situation and (2) discover whether the changes in mechanical properties align with the expected effect of changes in crystallinity.

\subsubsection{DSC/TGA}

Crystallinity is important for a number of reasons, many of which have been previously mentioned. Although the other tests may indicate changes in crystallinity, it is important to use DSC to accurately identify these changes. In this test, the properties of a material in response to temperature are analyzed \cite{Gill2010}, generating DSC plots that allow one to calculate the percent crystallinity of a material. As the degradation temperature of the materials was unknown, thermogravimetric analysis (TGA) was first conducted to determine an acceptable range of temperatures for DSC. TGA is similar to DSC but specifically measures the weight of a material in response to temperature \cite{Ebnesajjad2006}.

\subsubsection{Limitations of The Tests}

Another property that has been demonstrated to change as a result of recycling is the molecular weight of the material. Average molecular weight, which is essentially the average weight of each molecule in a material, significantly affects the mechanical properties of a material \cite{Gleadall2015}. In many cases, mechanical recycling reduces molecular weight \cite{Achukwu2023,Gleadall2015} which consequently causes a degradation in mechanical properties \cite{Raj2013,Zhou2023}. This was not specifically measured in this study.

\subsection{Specimen Preparation}

\subsubsection{Tensile and Charpy Tests}

In accordance with the standards, a type 1 tensile specimen with a 6mm thickness, 20mm width overall, and a 167mm length overall, as well as a Charpy specimen with a 10mm width, was designed using the Fusion 360 computer-aided design (CAD) software. Fusion 360 was chosen for familiarity with the software, accessibility, and quality. To design the specimens, the “sketch” tool was first used to draw a cross-section with the predefined variables and then the “extrude” tool was used to extrude the sketch to the correct thickness. Next, the tensile specimen thickness, the Charpy specimen height, and other dimensions were adjusted to account for shrinkage. These values can be found in Appendix~\ref{app:adj}.

After this, the design was exported as a mesh into a .stl file and imported into the online MakerBot Slicer to generate the print files. The tensile specimens were printed with their largest side facing up, while the Charpy specimens were printed with their notched side facing up. Unfortunately, the software did not have a PP-GF preset option, so one of the filament manufacturers was consulted and Nylon 12 Carbon Fiber was determined to be a possible substitute, which was then verified through testing. Other parameters were also adjusted to ensure the print quality for each material, which can be found in Appendix~\ref{app:param}.

Before printing, the bed was coated with Magigoo PP-GF adhesive to prevent warping. Subsequently, the spool of filament as well as the generated print file were loaded onto the Method X printer fitted with a LABs extruder, and the printing process began. There are a multitude of reasons one would favor this printer; however, for the purposes of this study, it was of interest due to its ease of access and excellent print quality, while the MakerBot slicer was solely chosen for its compatibility with the printer. Before Charpy testing, the specimens were notched with an 1mm depth. It is worth noting that these materials proved to be quite challenging to use, as it took numerous iterations of parameters to get them to print properly. It is possible that the specimens still had internal deformations from the printing process that caused deviations in mechanical properties, which were not considered in this study.
\subsubsection{DSC/TGA}

In TGA, each specimen was prepared by massing an Alumina 70~µl crucible and then inserting a small specimen cut from one of the filaments. For DSC analysis, the specimens were prepared by cutting off small pieces of filament with a mass of 5-10~mg. The specimens were first weighed, then placed in an Alumina 40~µl crucible, which was subsequently placed into the machine.

\subsection{Lab Testing}

\subsubsection{Tensile and Charpy Tests}

Once finished, the specimens were inspected for any deformities, placed in a secure container, and brought to the test site at UMASS Lowell. The two specimen types all sat at an approximate temperature of 68-70\textdegree F for 20+ hours before testing, except for those discussed in the Charpy results. At least five samples were tested for each material in both the Charpy and tensile tests. The tensile test was performed on an MTS Criterion Model 43 with a speed of 50.8~mm/min, a temperature of 71\textdegree F, and a humidity of 50\%. Charpy testing was conducted on an Instron Charpy impact machine with an 5~J capacity and a 229.7~mm span.

\subsubsection{DSC/TGA}

The DSC and TGA specimens were prepared in the lab, so testing began immediately after specimen preparation. TGA was first conducted on a Mettler Toledo TGA2 machine with a temperature range of 550\textdegree C, a heating rate of 10~k/min, and Nitrogen as the purge gas at a rate of 70~ml/min. Using this process, a 0-250\textdegree C temperature range was determined to be appropriate for DSC. DSC was then conducted on a Mettler Toledo DSC 3+ using Alumina 40~µl crucibles and 10~K/min as the heating and cooling rates. Nitrogen was used as the purge gas at a rate of 20~ml/min. As the test itself, the chamber was first heated to 250\textdegree C, held at this temperature for five minutes to remove the thermal history, and then cooled. This procedure was repeated for a total of two heat-cool cycles.


\subsection{Characterization}

\subsubsection{Tensile and Charpy Tests}

Using Microsoft Excel, stress-strain curves for the tensile tests of each material were analyzed in order to gain insight into the material's ductility, strength, and stiffness. The graphs, average Young’s modulus\footnote{The Young's modulus of a material is its resistance to changes in length while under tension lengthwise \cite{britannica}.}, and average ultimate tensile strength (the maximum stress before break) were then compared between the two materials, which can be found in the results section. Young’s Modulus is the initial slope of the stress strain curve, given by Eq.~(\ref{eqn:1}).

\begin{equation}\label{eqn:1}
    E=\frac{\sigma}{\epsilon}
\end{equation}

\noindent Where $E$ is Young’s modulus, in MPa, $\sigma$ is stress, in MPa, and $\epsilon$ is strain, in (mm/mm). If this value is high, the material has a high resistance to deformation. For the Charpy test, the average breaking energy for each material was compared. The standard deviation of each of these values was also calculated.

\subsubsection{DSC}

\begin{figure}[h]
\centering\includegraphics[width=.70\linewidth]{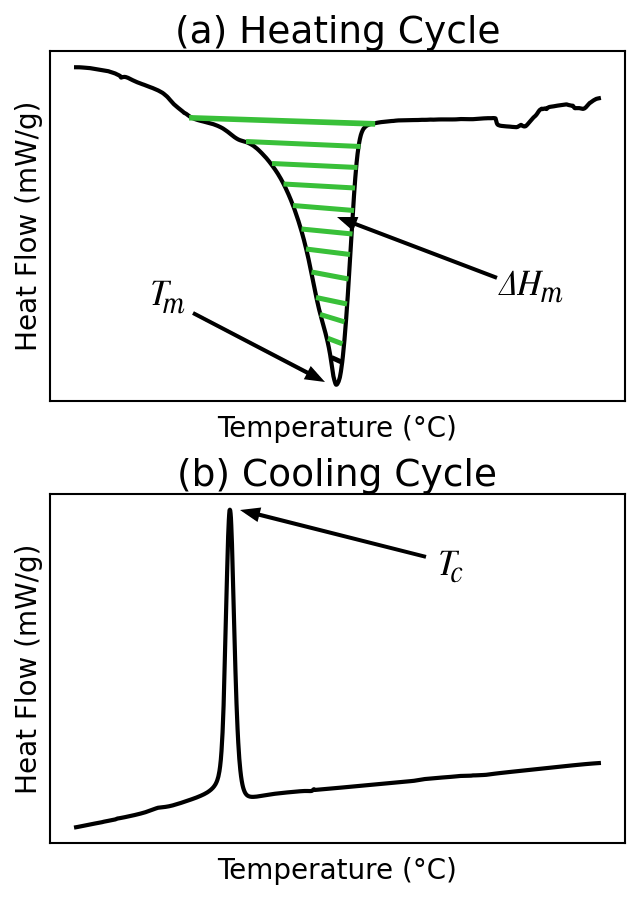}
\caption{A labeled example of DSC curve peaks resulting from the heating cycle (a) and the cooling cycle (b).\label{fig:2}}
\end{figure}

All calculations were performed based on the first cooling cycle and the second heating cycle, as the first heat was performed to remove the material’s thermal history. An example of the outcome of a heating and cooling cycle, in the form of a graph, is shown in Fig.~\ref{fig:2}.

\noindent Using the STARe Evaluation software to to find the area between the melting peak and the baseline (the dotted line) in the DSC curves, the heat of melting, $\Delta H_m$, was calculated. To calculate percent crystallinity, or $X_c$, Eq.~(\ref{eqn:2}) was used.

\begin{equation}\label{eqn:2}
    X_c=\frac{\Delta H_m}{(1-\omega_f) \Delta H_m^{\circ}}
\end{equation}

\noindent Where $\Delta H_m$ (in J/g) is the area between the baseline and the melting peak, $\Delta H_m^{\circ}$ is the theoretical melting enthalpy of PP (209J/g), and $\omega_f$ is GF percent in the material \cite{Yang2021}. This equation was adopted from a similar study authored by Chunxia Yang and others at Shandong University, which explored the mechanical and material properties of PP and PP-GF foams. The average crystallinity, melting temperature ($T_m$), and crystallization temperature ($T_c$) were calculated for each material, and then compared.

\section{Results}

\subsection{DSC}

\begin{figure}[h]
\centering\includegraphics[width=.75\linewidth]{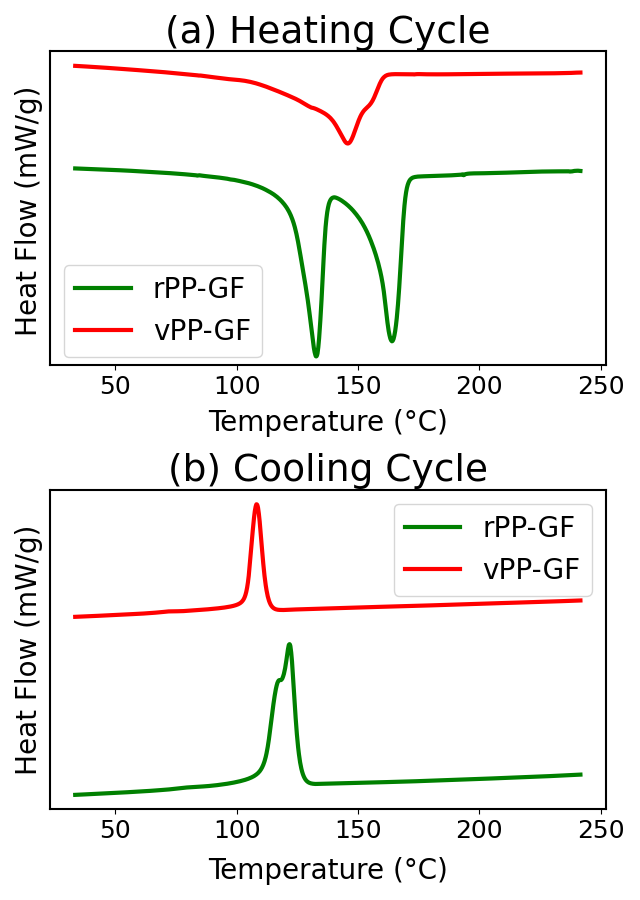}
\caption{The DSC thermographs for the heating cycle (a) and the cooling cycle (b) for rPP-GF and vPP-GF; vPP-GF curves are shifted vertically.\label{fig:3}}
\end{figure}

As GF has a much higher melting temperature than the DSC temperature range used, the thermographs in Fig.~\ref{fig:3} only capture the crystallization and melting data for the non-GF components of each material. For rPP-GF, this was initially assumed to be rPP, and for vPP-GF, it was assumed to be vPP. The melting thermograph for rPP-GF contains two melting peaks, which is common for composite materials. This is likely also the reason that rPP-GF has a wider and taller crystallization peak. In Table~\ref{tab:4}, it can be seen that vPP-GF has a higher crystallinity.

Theoretically, the heating thermograph for rPP-GF should only contain one peak (for rPP), so the fact that there are two indicates the presence of another material that is not reported by the manufacturer. Out of the two peaks, the latter is assumed to be rPP, as the peak temperature is similar to ones seen in other studies \cite{Yang2021, Zdiri2018, Aurrekoetxea2001}. Determining the material corresponding to the first peak poses a challenge because there is little known beyond its melting point. With this being said, revisiting the origins of rPP-GF yields a potential candidate: HDPE. This is promising for a number of reasons: (1) HDPE has a similar melting point to the one calculated \cite{Li2019} and (2) rPP-GF originates from fishnet and rope, which is primarily composed of PP and HDPE. Thus, HDPE is considered to be the unknown material for the remainder of this paper. If this was not a deliberate choice by the manufacturer, it may be that inadequate filtering led to residual HDPE in the final product.

\begin{table}[h]
\caption{DSC Results}\label{tab:4}
\centering
\begin{tabular}{!{\hspace*{0.5cm}} l @{\hspace*{1cm}} l  @{\hspace*{.7cm}} l  @{\hspace*{.7cm}} l @{\hspace*{.7cm}} l !{\hspace*{0.5cm}}}
\toprule
\textbf{Material} & 
\textbf{$T_m$ (\textdegree C)} & 
\textbf{$T_c$ (\textdegree C)} &
\textbf{$X_c$ (\%)} \\
\midrule 
\addlinespace[.2em]
vPP-GF & 145.6 & 108.4 & 26.4\\[.5em]
rPP-GF & 132.5/163.6 & 122.2 & 23.0\\
\addlinespace[.2em]
\bottomrule
\end{tabular}
\end{table}

Although the percent crystallinity of rPP initially appears to be lower than vPP, the calculation was done under the assumption that the percent of rPP in rPP-GF was exactly 72.5\%. Without knowing the wt\% of HDPE it is impossible to accurately calculate the wt\% of rPP and subsequently the percent crystallinity of rPP. Given the additional presence of HDPE, the wt\% of rPP must be lower than originally thought, so the actual $X_c$ for rPP must be higher. Furthermore, the two rPP heating peaks were closely spaced, causing interference between them. It was consequently challenging to establish an appropriate baseline to accurately calculate the entire melting peak area for rPP, which resulted in a lower calculated $X_c$. In light of these errors, it is not unlikely that the crystallinity of is close to, if not higher, than that of the vPP.

The crystallization of rPP and HDPE interfere during the cooling cycle, resulting in a larger crystallization peak. This makes it difficult to precisely calculate the rPP crystallization temperature. Though, the peak begins after the crystallization for vPP, so it can still be assumed to be higher.

Notably, the vPP has a lower $T_m$ and $T_c$ when compared with the rPP, and a significantly lower $X_c$, $T_m$, and $T_c$ when compared with other values in the literature \cite{Huang2021, Aurrekoetxea2001, Yang2021}. The disparity between the $T_m$ and $T_c$ for virgin and recycled specimens somewhat aligns with a study by Huang et al., where they find that recycling increases $X_c$, $T_m$, $T_c$, and reduces molecular weight \cite{Huang2021}. This would make sense if rPP-GF had a higher $X_c$, though it is unknown at this point if that is true. Furthermore, the substantial difference between $T_m$ and $T_c$ values for vPP and the literature/rPP means recycling is unlikely to be the sole cause. As the materials are created by different manufacterers, it may be the result of a difference in material sourcing; rPP could be closer to isotactic PP, while vPP could be closer to amorphous PP. This would also make sense if the rPP has a higher $X_c$.

\subsection{Tensile}

\begin{figure}[h]
\centering\includegraphics[width=.80\linewidth]{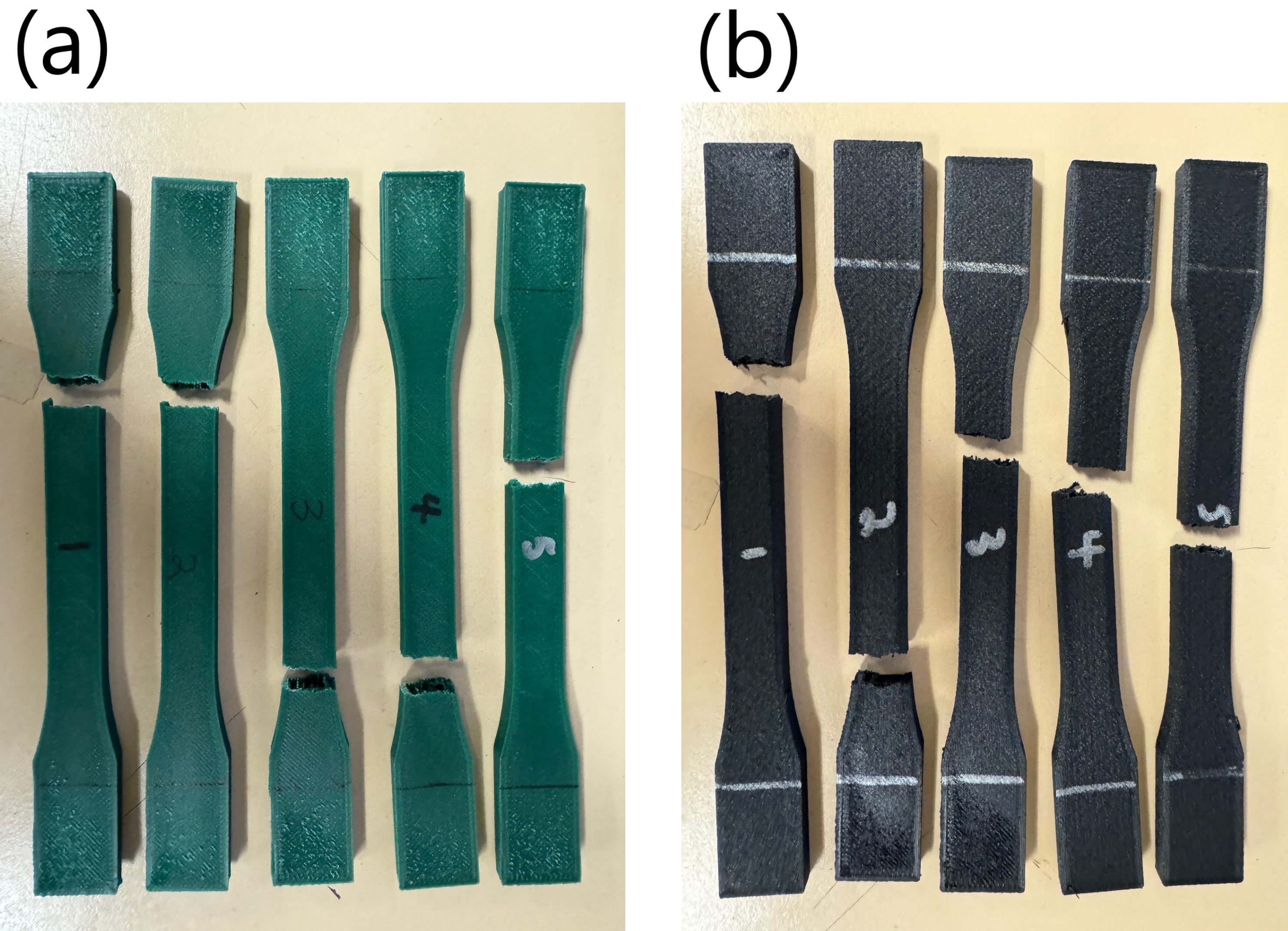}
\caption{A picture of the fractured rPP-GF (a) and vPP-GF (b) tensile specimens.\label{fig:4}}
\end{figure}

\begin{figure}[h]
\centering\includegraphics[width=.75\linewidth]{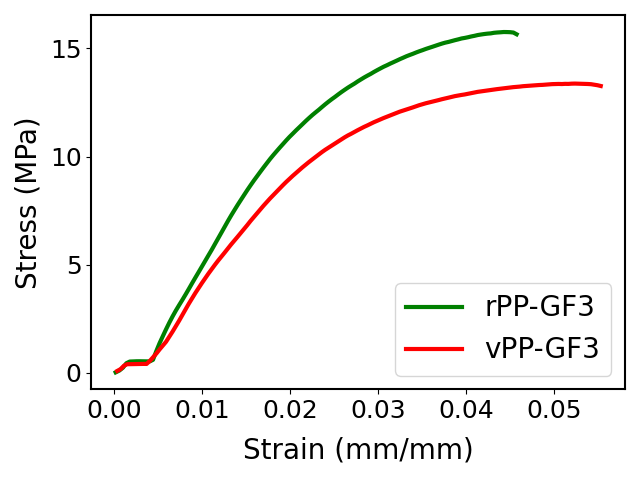}
\caption{Example stress-strain curves best that represent the dataset.\label{fig:5}}
\end{figure}

Figure~\ref{fig:4} displays a picture of the fractured rPP-GF and vPP-GF specimens, respectively. Figure~\ref{fig:5} shows stress-strain curves that accurately represent the results. It can be seen from these that both materials exhibit brittle fracture behavior. However, the materials differ when it comes to fracture location; rPP-GF consistently broke towards the end of the specimen, while vPP-GF was more varied in terms of fracture location.

\begin{figure}[h]
\centering\includegraphics[width=.75\linewidth]{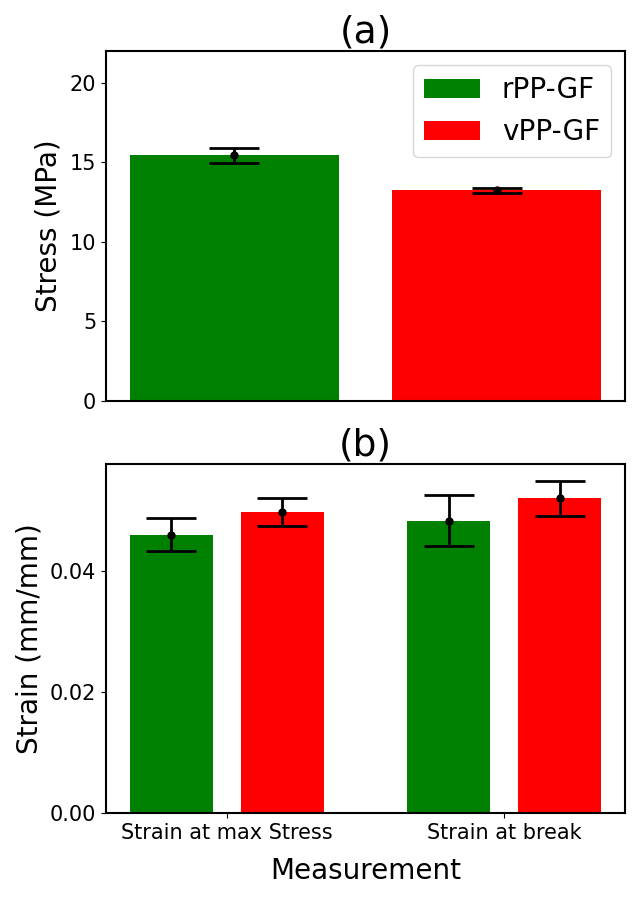}
\caption{Bar graphs comparing the max stress (a) as well as strain at max stress and strain at break (b) between rPP-GF and vPP-GF.\label{fig:6}}
\end{figure}

\begin{figure}[h]
\centering\includegraphics[width=.75\linewidth]{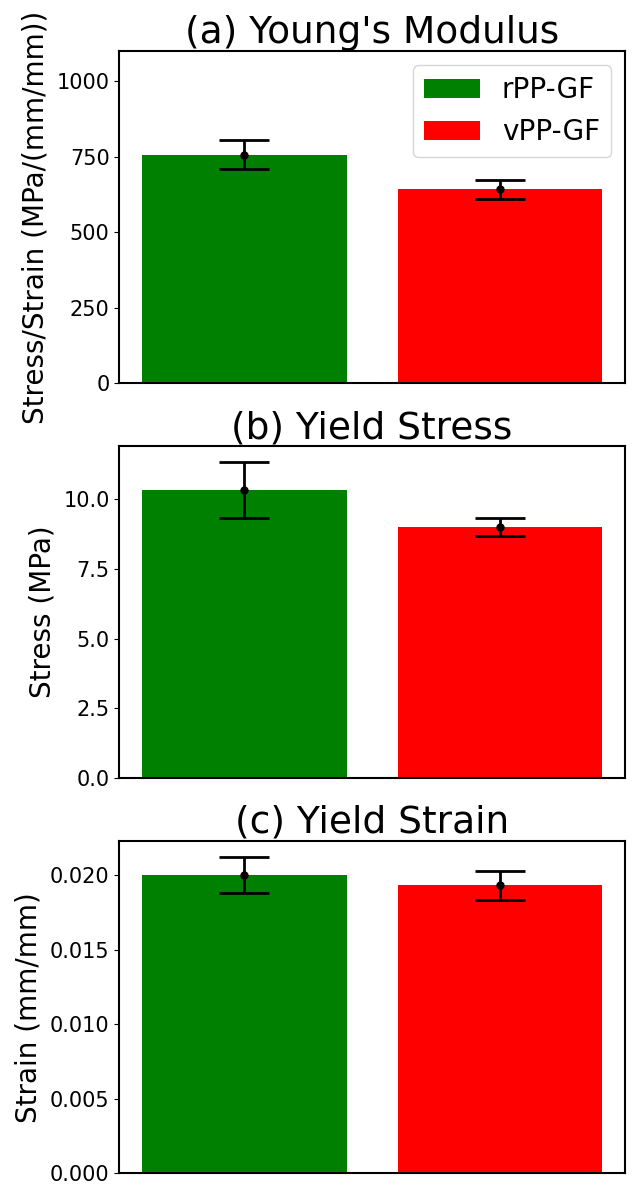}
\caption{Bar graphs comparing the Young's Modulus (a), Yield Stress (b), and Yield Strain (c) between rPP-GF and vPP-GF.\label{fig:7}}
\end{figure}

The max stress ($\sigma_{max}$), strain at max stress ($\epsilon_{ms}$), and strain at break ($\epsilon_b$) are compared in Fig.~\ref{fig:6}. It was found that rPP-GF had a higher $\sigma_{max}$, while vPP-GF had a higher $\epsilon_{ms}$ and $\epsilon_b$. For reasons previously discussed, the higher $\sigma_{max}$ can likely be attributed to a higher crystallinity. This would be consistent with what others have found when they saw an increase in max tensile stress \cite{Aurrekoetxea2001, Huang2021}. As for the lower $\epsilon_{ms}$ and $\epsilon_b$ observed in rPP-GF, it may be the result of a lower molecular weight. Recycled materials frequently exhibit a lower molecular weight, which can lead to a lower number of chains in the crystal to withstand stress, causing the material to fail at a lower elongation \cite{Aurrekoetxea2001}. Additionally, a lower $\epsilon_{ms}$ and $\epsilon_b$ suggest an increase in brittleness, something common in the literature \cite{Aurrekoetxea2001, Satya2020}.

The modulus, yield strength, and yield strain are compared in Fig.~\ref{fig:7}. It can be observed that rPP-GF exhibits a higher modulus, yield strength, and marginally higher yield strain. The higher modulus and yield strength are likely due to a higher crystallinity, while the difference in yield strain is nearly negligible. A crystalline material is more stiff, which is why a higher crystallinity can lead to a higher modulus and higher strain at yield \cite{Aurrekoetxea2001}.

It is worth noting that rPP-GF warped significantly more than vPP-GF. As HDPE is known for its high tendency to warp, this may be a further indication of the presence of HDPE within rPP-GF. The increase in strength contradicts the initial prediction that rPP-GF would exhibit degraded mechanical properties in all aspects. Although it is difficult to pinpoint an exact cause, the most obvious answer would be that a difference in material sourcing between the manufacturers led to the discrepancy. This could also indicate that recycling is much more dependent on the source of the actual material, not just whether the initial material is virgin or recycled. Moreover, it is unlikely that HDPE was the cause, as both PP and HDPE exhibit similar tensile properties \cite{Awad2019}. If anything, the presence of HDPE would cause a lower tensile strength, not a higher one.

Additionally, it can be seen that the standard deviation for rPP-GF values was consistently higher than for vPP-GF. The properties of plastic products (like fishnet and rope) can vary significantly depending on their source and manufacturing process, so it may be that using recycled material to make filaments causes more variance in the filament's properties.

\section{Charpy}

\begin{figure}[h]
\centering\includegraphics[width=.75\linewidth]{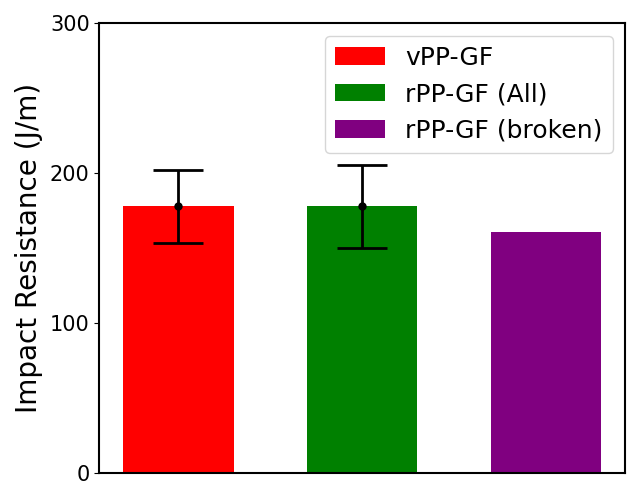}
\caption{Bar graphs comparing average impact resistance for vPP-GF, rPP-GF (including non-broken specimens), and rPP-GF (excluding broken specimens).\label{fig:9}}
\end{figure}

Unfortunately, three of the five Charpy specimens tested did not fully break into two separate pieces, which means that their results are considered invalid. A picture showcasing this is shown in Fig.~\ref{fig:9}. Furthermore, some vPP-GF specimens were not able to undergo the proper conditioning period because of time constraints, which could have led to some specimens being warmer than expected, causing a higher observed Charpy impact strength. These same constraints also meant that no additional specimens could be tested. With this being said, the results were still compared to attempt to glean the most information possible.

\begin{figure}[h]
\centering\includegraphics[width=.75\linewidth]{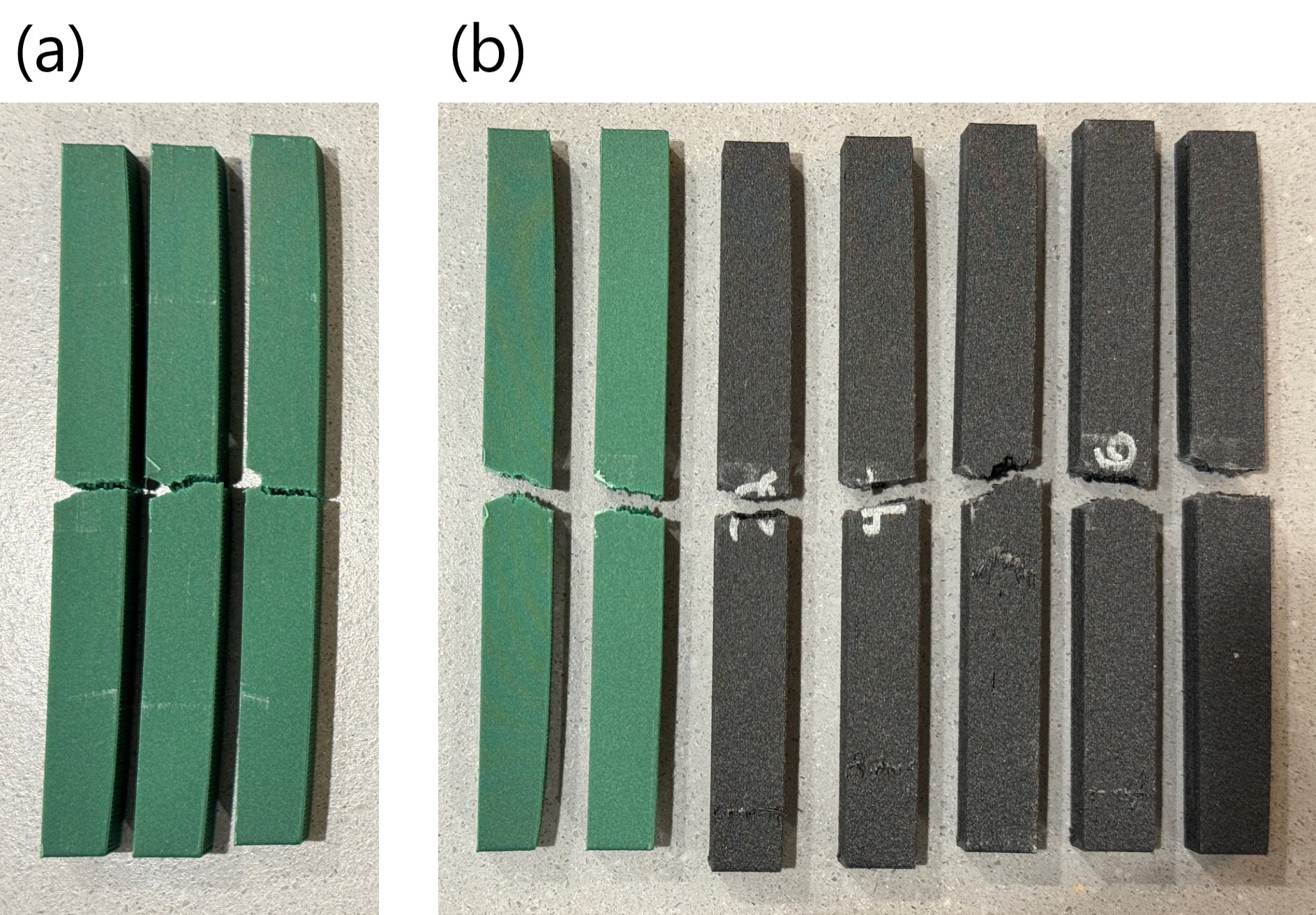}
\caption{Picture displaying the fractures for the unbroken specimens (a) and the broken specimens (b). The green specimens are rPP-GF, while the black specimens are vPP-GF.\label{fig:8}}
\end{figure}

From Fig.~\ref{fig:8}, it can be determined that both materials experienced brittle fracture. Figure~\ref{fig:9} reveals that the average impact resistance for rPP-GF (including non-broken specimens) was marginally higher than vPP-GF, while the average impact resistance (excluding the non-broken specimens) was lower than vPP-GF. The impact resistance for rPP-GF was remarkably closer to vPP-GF than expected. This can likely be attributed to the presence of HDPE within the rPP-GF composite; in one particular study on the mechanical properties of PP and HDPE, they found that the impact strength for HDPE was nearly four times that of PP \cite{Awad2019}. As a result, it is unknown how recycling affects the impact strength of rPP. The average impact strength for the broken rPP-GF specimens was not compared with vPP-GF as there were only two data points, which was insufficient.

\section{Limitations/Future Work}

Aside from the limitations already mentioned in this paper, the most concerning limitation has to do with the effect of recycling PP-GF. It has been documented that the glass fibers in PP-GF degrade from reprocessing, which causes a degradation of mechanical, thermal, physical, and rheological properties \cite{Shuib2023}. Essentially, recycling PP to form rPP-GF composites is only a temporary solution that extends the life span of PP. Potentially, it could still serve as a placeholder while more research is done on ways to recycle and re-use plastic. Additionally, limitations in terms of the materials used made it difficult to compare the two materials in some instances. In the future, a much clearer understanding would be gained from taking a single batch of PP, subjecting it to multiple reprocessing cycles to obtain rPP with varying degrees of reprocessing, and then forming rPP-GF composites corresponding to reprocessing each level to analyze.

\section{Conclusion}

In this research, two commercially available glass fiber-reinforced polypropylene composites (rPP-GF and vPP-GF) were compared through differential scanning calorimetry (DSC), tensile, and impact testing. One was purely virgin, and the other was a combination of rPP (from fishnets and rope) and vGF. They were analyzed to determine the effect of recycling PP and reinforcing it with a glass fiber as a potential solution to extend the lifetime of PP and reduce plastic buildup in the oceans. From the research, it can be concluded that:

\begin{enumerate}
    \item[(1)] The rPP-GF composite analyzed had another material present that was not reported by the manufacturer. This was determined to likely be HDPE. It was also found that rPP-GF had a higher melting peak and crystallization peak temperature, and in light of errors/limitations in terms of calculations, it likely also had a higher crystallinity.
    \item[(2)] rPP-GF outperformed vPP-GF in terms of, max stress, yield stress, and Young’s modulus, while vPP-GF outperformed rPP-GF in terms of strain at break and yield strain. Additionally, it was determined that rPP-GF was more brittle. These were taken as further indications that rPP-GF had a higher crystallinity than vPP-GF. The higher standard deviation for rPP-GF suggests that recycling fishnet and rope leads to more variance in properties.
    \item[(3)] The presence of HDPE in rPP-GF caused a higher observed impact strength, which made it difficult to discern how recycling affects rPP and combining it with vGF affects the impact strength.
\end{enumerate}

Ultimately, the fact that rPP-GF outperformed vPP-GF in multiple areas is strong evidence for the potential of recycling polypropylene fishnet/rope and reinforcing it with glass fibers to extend their lifespan and reduce ocean plastic.

\section{Acknowledgements}

I thank Dr. Christopher J. Hansen (UMASS Lowell), for providing advice throughout the research process and coordinating testing; I thank Mr. David Steeves (Chelmsford High School) for the privilege of using the 3D printers and for the help printing the specimens; I thank Aidan Kenawell, Kyeuongbin Min, and Mr. Barrett O'Brien (UMASS Lowell) for help testing; and I thank Mr. Jonathan Morris, Mr. Axel Martinez, and Mr. Christopher DiCarlo (Chelmsford High School) for useful discussions regarding my research topic and paper format.


\appendix

\section{Specimen Adjustments}\label{app:adj}

In Table~\ref{tab:5}, the (height) column denotes \% increase for either the Charpy specimen height or the tensile specimen thickness, while the (other) column denotes the \% increase for all other specimen dimensions.

\begin{table}[h]
\caption{Specimen Adjustments}\label{tab:5}
\centering
\begin{tabular}{!{\hspace*{0.5cm}} l @{\hspace*{1cm}} l  @{\hspace*{.7cm}} l  @{\hspace*{.7cm}} l @{\hspace*{.7cm}} l !{\hspace*{0.5cm}}}
\toprule
\textbf{Material} & 
\textbf{Type} & 
\textbf{\% Inc} &
\textbf{\% Inc}\\
\textbf{} & 
\textbf{} & 
\textbf{(other)} &
\textbf{(height)}\\
\midrule 
\addlinespace[.2em]
vPP-GF & Tensile & -2.40 & 0.50\\[.5em]
rPP-GF & Tensile & -2.40 & 0.50\\[.5em]
vPP-GF & Charpy & 1.10 & 0.47\\[.5em]
rPP-GF & Charpy & -0.44 & 0.00\\
\addlinespace[.2em]
\bottomrule
\end{tabular}
\end{table}

\section{Printing Parameters}\label{app:param}

All parameters not specified in the rPP-GF table can be assumed to be the same as the vPP-GF table.

\begin{table}[h]
\caption{Specimen Printing Parameters (vPP-GF)}\label{tab:6}%
\centering{%
\begin{tabular}{!{\hspace*{0.5cm}}  p{3.5cm} @{\hspace*{1cm}} l !{\hspace*{0.5cm}}}
\toprule
\textbf{Parameter} & \textbf{Value} \\
\midrule
\addlinespace[.5em]
Extruder Temperature (\textdegree C)  & 240 \\[.5em]
Platform Temperature (\textdegree C) & 95\\[.5em]
Chamber Temperature (\textdegree C) & 38\\[.5em]
Layer Height (mm) & 0.25\\[.5em]
Retraction Speed (mm/s) & 10\\[.5em]
Retraction Distance (mm)  & 0.75\\[.5em]
Outline Print Speed (mm/s)  & 15\\[.5em]
Inset Print Speed (mm/s) & 15\\[.5em]
Solid Print Speed (mm/s) & 40\\[.5em]
Infill &  Hexagonal\\[.5em]
Infill Density (\%) &  50\\[.5em]
Number of Outer Walls & 4\\[.5em]
Active Cooling Layer & 4\\[.5em]
Raft & Yes\\ [.5em]
\bottomrule
\end{tabular}
}%
\end{table}

\begin{table}[h]
\caption{Specimen Printing Parameters (rPP-GF)}\label{tab:7}%
\centering{%
\begin{tabular}{!{\hspace*{0.5cm}}  p{3.5cm} @{\hspace*{1cm}} l !{\hspace*{0.5cm}}}
\toprule
\textbf{Parameter} & \textbf{Value} \\
\midrule
\addlinespace[.5em]
Chamber Temperature (\textdegree C) &  60\\[.5em]
Raft Top Layer \newline Speed (mm/s)  &   30\\[1.5em]
Raft Interface \newline Speed (mm/s) &  20\\[1.5em]
Use Active Cooling &  No\\[.5em]
\bottomrule
\end{tabular}
}%
\end{table}

\vfill\null


\bibliographystyle{asmejour}

\bibliography{citations_2}


\end{document}